\documentclass[conference]{IEEEtran}
\usepackage[pdftex]{graphicx}
\usepackage{caption}
\usepackage{subcaption}
\ifCLASSINFOpdf
\else
\fi
\usepackage{amsmath}
\usepackage{amssymb}

\newcommand{\argmin}{\operatornamewithlimits{argmin}}

\newcommand{\minimize}{\operatornamewithlimits{minimize}}

\newcommand{\st}{\operatornamewithlimits{\text{subject to}}}

\usepackage{algorithm} 
\usepackage{algorithmic}

\begin{document}
%
\title{DIeSEL: DIstributed SElf-Localization of a network of underwater vehicles}

\author{\IEEEauthorblockN{Cl\'{a}udia Soares, Pusheng Ji, Jo\~{a}o
    Gomes, and Antonio Pascoal} \IEEEauthorblockA{Institute for
    Systems and Robotics (ISR/IST) LARSyS, Instituto Superior Técnico,
    Universidade de Lisboa, Portugal\\
    Email:\{csoares,jpg,antonio\}@isr.ist.utl.pt,
    pusheng.ji@tecnico.ulisboa.pt}}


%


\maketitle

\begin{abstract}
  How can teams of artificial agents localize and position themselves
  in GPS-denied environments? How can each agent determine its
  position from pairwise ranges, own velocity, and limited interaction
  with neighbors? This paper addresses this problem from an
  optimization point of view: we directly optimize the nonconvex
  maximum-likelihood estimator in the presence of range measurements
  contaminated with Gaussian noise, and we obtain a provably
  convergent, accurate and distributed positioning algorithm that
  outperforms the extended Kalman filter, a standard centralized solution
  for this problem.
\end{abstract}


%
\IEEEpeerreviewmaketitle

\section{Introduction}
The deployment of networked systems of agents that can interact with
the physical world and carry out complex tasks in heterogeneous
environments is currently a major driver for research and
technological development. This trend is also seen in contemporary
ocean applications and it propelled research projects on multi-vehicle
systems like WiMUST (Al-Khatib et al.~\cite{al2015widely}).  Our work
concerns localization of (underwater) vehicle formations, a key
subsystem needed in the absence of GPS to properly georeference any
acquired data and also used in cooperative control algorithms.

\paragraph*{Related work}

The signal processing and control communities studied the network
localization problem in many variants, like static or dynamic network
localization, centralized or distributed computations,
maximum-likelihood methods, approximation algorithms, or outlier
robust methods.  

The control community's mainstream approach to localization relies on
the robust and strong properties of the Kalman filter to dynamically
compensate noise and bias. Recent approaches can be found in Pinheiro
et al.~\cite{PinheiroMorenoSousaRodriguez2016} and Rad et
al.~\cite{RadWaterschootToonLeus2011}. In the first very recent paper,
position and velocity are estimated from ranges, accelerometer
readings and gyroscope measurements with an Extended Kalman
filter. The authors of the second paper linearize the range-only
dynamic network localization problem, solving it with a linear Kalman
filter.

Our optimization-based approach is mainly inspired by the literature
on sensor network localization, or static \emph{cooperative}
localization, using pairwise range measurements.  Shang et
al.~\cite{ShangRumiZhangFromherz2004} follow a multidimensional
scaling approach, but multidimensional scaling works well only in
networks with high connectivity --- a property not encountered in
practice in large-scale geometric networks. Biswas et
al.~\cite{BiswasLiangTohYeWang2006} and O\u{g}uz-Ekim et
al.~\cite{OguzGomesXavierOliveira2011} proposed semi-definite and
second order cone relaxations of the maximum likelihood
estimator. Although more precise, these convexified problems get
intractable even for a small number of nodes.  The methods in
Calafiore et al.~\cite{CalafioreCarloneWei2010} and Soares et
al.~\cite{SoaresXavierGomes2014a} increase the precision of an initial
good guess, but are prone to local minima if wrongly initialized.  To
disambiguate the spatial configuration, classic static range-based
localization requires a minimum of three non-collinear anchors (i.e.,
reference points with known position) for a planar network and four
anchors for a volumetric one. In practice, the number of anchors might
be larger to attain the desired positioning accuracy.  Lately, signal
processing researchers produced solutions for dynamic network
localization; Schlupkothen et
al.~\cite{SchlupkothenDartmannAscheid2015} and Soares et
al.~\cite{SoaresGomesFerreiraCosteira2017} incorporated velocity
information from past position estimates to bias the solution of a
static relaxation of the localization problem via a regularization
term.

In our work we exploit the fact that underwater vehicles are
sophisticated nodes that often provide additional measurements besides
ranges, such as depth (from pressure sensors) and relative velocity (from
DVL). We leverage this side information to reduce the required number
of anchors, which might be cumbersome or costly to deploy in
underwater applications --- for example, when anchors are GPS-enabled
surface vehicles every reduction affords substantial savings in terms
of logistics and budget.

\paragraph*{Our approach}
\label{sec:our-approach}
We adopt a data model for cooperative localization of underwater
vehicle formations consisting of a modest number, or even a single
anchor and measurements for ranges, depths, and vehicle velocities
relative to the water, whenever these quantities are available. We
formulate the problem as a maximum-likelihood estimation problem
assuming noise on the range measurements. We derive a distributed
algorithm to provably attain a stationary point of the nonconvex
maximum-likelihood estimator.

\section{Problem formulation}
\label{sec:problem-formulation}

We denote the measurement and communication graph connecting vehicles
as~$\mathcal{G} = (\mathcal{V},\mathcal{E})$, where the node
set~$\mathcal{V} = \{1,2,\cdots,n\}$ represents the ensemble of
vehicles with unknown positions. Edge~$i \sim j \in \mathcal{E}$ means
that there is a range measurement and a communication link between
nodes~$i$ and~$j$. The set of vehicles with known positions, also
called anchors, is denoted~$\mathcal{A}= \{1, \cdots, m\}$ and we
assume~$m$ small (as low as~$m=1$). For each~$i \in \mathcal{V}$, we
let~$\mathcal{A}_{i} \in \mathcal{A}$ be the subset of anchors (if
any) relative to which node~$i$ also possesses a noisy range
measurement. To simplify notation, and without loss of generality, we
omit the explicit dependency in time of graph~$\mathcal{G}$.

Let~$\mathbb{R}^{d}$ be the space of interest, where~$d=2$ whenever
the vehicles move in a planar configuration or when we have access to
the depth of all vehicles, and~$d=3$ otherwise,
$p_{i} \in \mathbb{R}^{d}$ is the position of vehicle~$i$ at some
discrete time instant~$t-T_{0}$, with~$T_{0}$ as a convenient time
window,~$v^{I}_{i}(t) = v^{R}_{i}(t) + v_{f}$ is the inertial velocity at
the discrete time instant~$t$, where~$v^{R}_{i}(t)$ is vehicle's~$i$
known velocity relative to the fluid, and~$v_{f} \in \mathbb{R}^{d}$
is the unknown fluid velocity (current), assumed constant in space and
time. We define~$v_{i}(t) = \sum_{\tau =
  t-T_{0}}^{t}v_{i}^{R}(\tau)$. Thus, the position of node~$i$ at a
time~$t$
is~$x_{i}(t) = p_{i} + \sum_{\tau = t-T_{0}}^{t}v_{i}^{I}(\tau) \Delta
T$, where~$\Delta T$ is the sampling interval.  Let~$d_{ij}(t)$ be the
noisy range measurement between vehicles~$i$ and~$j$ at time~$t$,
known by both~$i$ and~$j$. Without loss of generality, we
assume~$d_{ij}(t) = d_{ji}(t)$. Anchor positions are denoted
by~$a_{k}(t) \in \mathbb{R}^{d}$. Similarly,~$r_{ik}(t)$ is the noisy
range measurement between node~$i$ and anchor~$k$.
Define~$x(t) = \{x_{i}(t) : i \in \mathcal{V}\}$ as the concatenation
of all unknown vehicle positions at time~$t$.

The distributed network localization problem addressed in this work
consists in estimating the set of vehicles' positions~$x(t)$, from the
available measurements
$\{d_{ij}(t) : i \sim j\} \cup \{r_{ik}(t) : i \in \mathcal{V}, k \in
\mathcal{A}_{i}\}$, through cooperative message passing between
neighboring sensors in the communication graph~$\mathcal{G}$.  Under
the assumption of zero-mean, independent and identically-distributed,
additive Gaussian measurement noise, the maximum likelihood estimator
for the sensor positions at time~$t$ is the solution of the
optimization problem
\begin{equation}
  \label{eq:orig-problem}
 \minimize  f_{T_{0},t}(x(t)),
 \end{equation}
where the cost is defined as
\begin{equation}
  \label{eq:cost1}
  \begin{split}
  f_{T_{0},t}(x(t)) =& \sum_{\tau = t-T_{0}}^{t} \left( \sum_{i \sim j}
    \frac 12 (\|x_{i}(\tau) - x_{j}(\tau)\|-d_{ij}(\tau))^{2} +
                        \right. \\
                    &\left. \sum_{i} \sum_{k \in \mathcal{A}_{i}} \frac 12
               (\|x_{i}(\tau) - a_{k}(\tau)\|-r_{ik}(\tau))^{2} \right).
  \end{split}
\end{equation}

\section{Underwater localization}
\label{sec:underwater-local}

Define~$x_{i} = \{x_{i}(\tau)\}$ and introduce two sets of auxiliary
variables~$y_{ij} = \{y_{ij}(\tau)\}$ and~$w_{ik} =
\{w_{ik}(\tau)\}$. Adopting the approach of~\cite{SoaresXavierGomes2014a}
we can reformulate problem~\eqref{eq:orig-problem} as
\begin{equation}
  \label{eq:reformulated1}
  \begin{split}
     \minimize_{\{x_{i}\}, \{y_{ij}\}, \{w_{ik}\}} & \sum_{i \sim j} \sum_{\tau = t-T_{0}}^{t} \frac
                12 \|x_{i}(\tau) - x_{j}(\tau) - y_{ij}(\tau)\|^{2} 
                         +\\
              & \sum_{i} \sum_{k \in \mathcal{A}_{i}} 
                \sum_{\tau = t - T_{0}}^{t} \frac 12
               \|x_{i}(\tau) - a_{k}(\tau) -w_{ik}(\tau)\|^{2}
                \\
    \st & \|y_{ij}(\tau)\| = d_{ij}(\tau), \; \forall i \sim j\\
        & \|w_{ik}(\tau)\| = r_{ik}(\tau), \; \forall i \in \mathcal{V}, k \in \mathcal{A}_{i}.
  \end{split}
\end{equation}
We now notice that the subtraction~$x_{i}(\tau) - x_{j}(\tau)$ cancels
out the fluid velocity, obtaining, for both node-node and node-anchor pairs,
\begin{equation*}
  \begin{split}
    x_{i}(\tau) - x_{j}(\tau) &= p_{i} - p_{j} + (v_{i}(\tau) -
    v_{j}(\tau))\Delta T\\
    x_{i}(\tau) - a_{k}(\tau) &= p_{i} - q_{k} + (v_{i}(\tau) -
    u_{k}(\tau))\Delta T.
  \end{split}
\end{equation*}
where the anchor decomposed
as~$a_{k}(\tau) = q_{k} + u_{k}(\tau) \Delta T$ mirrors the
aforementioned decomposition for nodes. Define the
concatenation~$z = (\{p_{i}\}, \{y_{ij}\}, \{w_{ik}\})$,
where~$y_{ij} = \{y_{ij}(\tau)\}$ and~$w_{ik} = \{w_{ik}(\tau)\}$
stack~$y_{ij}(\tau)$ and~$w_{ik}(\tau)$ for all time instants
from~$\tau = t-T_{0}$ to~$\tau = t$, and the constraint set
\begin{equation}
  \label{eq:constraint-set}
  \mathcal{Z} = \{ z : \|y_{ij}\| = d_{ij} \forall i\sim j, \|w_{ik}\|
  = r_{ik} \forall i \in \mathcal{V},  k \in \mathcal{A}_{i}\}.
\end{equation}
The terms in~$\tau$ in the cost of problem~\eqref{eq:reformulated1}
can be stacked, thus obtaining the problem
\begin{equation}
  \label{eq:time-stacked}
  \begin{array}{rl}
    \minimize & \frac 12 \|DAp + \Delta v - y \|^{2} + \frac 12 \|Ep
                -\alpha - w \|^{2},\\
    \st & (p,y,w) \in \mathcal{Z}
  \end{array}
\end{equation}
where we denote matrix~$A$ as the Kronecker product of the arc-node
incidence matrix~$C$ of the measurement graph~$\mathcal{G}$ with the
identity matrix~$I_{d}$, matrix~$D$ is a tall concatenation of
identity matrices and matrix~$E$ is a selector with zeros and
ones. The constant~$\Delta v$ is defined as
\begin{equation*}
  \Delta v = \{(v_{i}(\tau) - v_{j}(\tau)) \Delta T, \text{ for all } \; i \sim j,
t - T_{0} \leq \tau \leq t\},
\end{equation*}
and the anchor term's non-optimized data term is
\begin{equation*}
  \begin{split}
    \alpha = &\{ q_{k} - (v_{i}(\tau)-u_{k}(\tau))\Delta T, \\
    &\text{ for
      all } \; i \in \mathcal{V}, k \in \mathcal{A}_{i}, t - T_{0}
    \geq \tau \geq t\}.
  \end{split}
\end{equation*} 
With this notation, Problem~\eqref{eq:time-stacked} is equivalent to
\begin{equation*}
  \begin{array}{rl}
    \minimize & \frac 12 \left \|
                \begin{bmatrix}
                  DA & -I & 0
                \end{bmatrix}
                            \begin{bmatrix}
                              p\\y\\w
                            \end{bmatrix} + \Delta v
    \right \| ^{2} +\\
& \frac 12 \left \|
                \begin{bmatrix}
                  E & 0 & -I
                \end{bmatrix}
                            \begin{bmatrix}
                              p\\y\\w
                            \end{bmatrix}
    - \alpha \right \| ^{2}\\
    \st & (p,y,z) \in \mathcal{Z}.
  \end{array}
\end{equation*}
The cost can be further simplified as
\begin{equation}
  \label{eq:quadratic-cost}
  \begin{array}{rl}
  \minimize & F_{T_{0},t}(z) = \frac 12 z^{T}Mz - b^{T}z\\
    \st & z \in \mathcal{Z}
  \end{array}
\end{equation}
where matrix~$M$ is
\begin{equation*}
  M =
  \begin{bmatrix}
    A^{T}D^{T}DA + E^{T}E & -A^{T}D^{T} & -E^{T}\\
    -DA & I & 0 \\
    -E & 0 & I
  \end{bmatrix},
\end{equation*}
and vector~$b$ is
\begin{equation*}
  b =
  \begin{bmatrix}
    E^{T}\\0\\-I
  \end{bmatrix}
\alpha -
\begin{bmatrix}
  A^{T}D^{T} \\ -I \\ 0
\end{bmatrix}
\Delta v.
\end{equation*}
The cost in problem~\eqref{eq:quadratic-cost} is quadratic, but as its
minimization is constrained to a nonconvex set it retains the hardness
of the original, equivalent, problem~\eqref{eq:orig-problem}. Also,
off-diagonal terms in matrix~$M$ couple variables for the different
vehicles, which impedes a distributed solution. In the next section we
will approach these issues to achieve a distributed positioning
algorithm with provable convergence.

\section{Distributed underwater localization}
\label{sec:distr-underw-local}


Problem~\eqref{eq:quadratic-cost} is a quadratic cost constrained to a
nonconvex set, which is a very hard optimization problem, in
general. Further, the off-diagonal terms couple the vehicles'
variables, thus preventing a distributed solution. Nevertheless, we
are now, at each time step, in the conditions of the algorithm
presented in~\cite{SoaresXavierGomes2014a}. From here on we adapt the
results of this reference to the current problem.

We now call the reader's attention for the fact that every quadratic
function has a Lipschitz continuous gradient, meaning that we can
\emph{majorize} our correlated quadratic with a diagonal scale term,
the Lipschitz constant~$L$ associated with the function's gradient
\begin{equation*}
  \begin{split}
    F_{T_{0},t}(z) \leq & F_{T_{0},t}(z(\kappa)) + \langle \nabla F_{T_{0},t} (z(\kappa)),
    z-z(\kappa)\rangle + \\ & \frac L2 \|z-z(\kappa)\|^{2},
  \end{split}
\end{equation*}
valid for any points~$z,$ and~$z(k)$.  We iteratively majorize the
quadratic coupled cost by this quadratic diagonal upper bound of the
cost, and then minimize the majorizer, under the framework of a
Majorization-Minimization algorithm. Here the algorithm steps are
denoted by~$\kappa$, to avoid confusion with the discrete time running
on~$t$. Minimization of the majorizer can be written as
\begin{equation}
  \label{eq:descent}
  \begin{split}
    z(\kappa+1) = \argmin_{z \in \mathcal{Z}} & F_{T_{0},t}(z(\kappa)) + \langle
    \nabla F_{T_{0},t} (z(\kappa)), z-z(\kappa)\rangle +\\
    & + \frac L2 \|z-z(\kappa)\|^{2}.
  \end{split}
\end{equation}
The solution to Problem~\eqref{eq:descent} is the well-known projected
gradient iteration
\begin{equation}
  \label{eq:projected-gradient}
  z(\kappa+1) = \mathrm{P}_{\mathcal{Z}}\left( z(\kappa) - \frac1L \nabla
    F_{T_{0},t}(z(\kappa)) \right),
\end{equation}
where~$\mathrm{P}_{\mathcal{Z}}(z)$ is the projection of point~$z$
onto set~$\mathcal{Z}$. The gradient~$\nabla F_{T_{0},t}(z)$ is
simply
\begin{equation*}
  \nabla F_{T_{0},t}(z)= Mz-b,
\end{equation*}
and a Lipschitz constant~$L$ associated with the
gradient~$\nabla F_{T_{0},t}(z)$ to compute the gradient step
in~\eqref{eq:projected-gradient}
turns out to be~$L = T_{0}(2\delta_{\mathrm{max}} + \max_{i}|\mathcal{A}_{i}|) +
2$, where~$\delta_{\mathrm{max}}$ is the maximum node degree in the
network.
\begin{algorithm}[tb]
  \caption{DIeSEL: Distributed Self-Localization}
  \label{alg:DIeSEL}
  \begin{algorithmic}[1] 
    \REQUIRE~~\\
    $L;$ \\
    $ z = (p,y,w);$\\
    $\beta_{i} = \frac{L - T_{0}(\delta_{i}-|\mathcal{A}_{i}|)}{L};$ \\
    $\{d_{ij}(t) : i \sim j \in \mathcal{E}\};$ \COMMENT{acquired as time~$t$ evolves}\\
    $\{r_{ik}(t) : i \in \mathcal{V}, k \in \mathcal{A}\};$ \COMMENT{acquired as time~$t$ evolves}
    \ENSURE $\hat{z} = \{(p_{i}, y_{ij}, w_{ik})\}$
    \FORALL{$t \in \{T_{0}, ...\}$}
    \STATE $\kappa = 1$\\
    $\Delta v_{ij} = \{ (v_{i}(\tau) -v_{j}(\tau)) \Delta T, \text{
      for all } t-T_{0} \leq \tau \leq t\}$ \\
    $\Delta u_{ik} = \{ (v_{i}(\tau) -u_{k}(\tau)) \Delta T, \text{
      for all } t-T_{0} \leq \tau \leq t\}$
    \WHILE{some stopping criterion is not met, each node $i$}
    \STATE broadcast $p_{i}$ to all neighbors
    \STATE 
    \begin{equation*}
      \begin{split}
        p_i^{+} =& \beta_{i} p_i + \\ &\sum_{j \in \mathcal{N}_{i}}
        \sum_{\tau = t-T_{0}}^{t}\left(p_j + C_{(i \sim j, i)} y_{ij}(\tau) +
          \Delta v_{ij}(\tau) \right) + \\ &\sum_{k \in |\mathcal{A}_i|}
        \sum_{\tau = t-T_{0}}^{t} \left( a_k + w_{ik}(\tau) + \Delta
          u_{ik}(\tau) \right)
      \end{split}
    \end{equation*}
    \FORALL{ $j$, $i \sim j$, and $\tau \in [t-T_{0}; t]$}
    \STATE
    \begin{equation*}
      \begin{split}
        y_{ij}^{+}(\tau) = &\mathrm{P}_{\mathcal{Y}_{ij}(\tau)} \Big(
        \frac{L-1}L y_{ij}(\tau) + \\
        &\frac 1L C_{(i \sim j,i)} (p_{i} -
        p_{j} - \Delta v_{ij}(\tau))\Big)
      \end{split}
    \end{equation*}
    \ENDFOR
    \FORALL{ $k \in \mathcal{A}_i$, and $\tau \in [t-T_{0}; t]$}
    \STATE
    \begin{equation*}
      w_{ik}^{+}(\tau) = \mathrm{P}_{\mathcal{W}_{ik}(\tau)}\left( \frac{L-1}L
        w_{ik}(\tau) + \frac 1L (p_{i} -a_{k} - \Delta u_{ik}(\tau))\right)
    \end{equation*}
    \ENDFOR
    \STATE update $\kappa = \kappa+1;$ $(p_{i}, y_{ij}, w_{ik}) = (p_{i}^{+}, y_{ij}^{+}, w_{ik}^{+})$
      
  %
    \ENDWHILE
    \ENDFOR
  \end{algorithmic}
\end{algorithm}
Algorithm~\ref{alg:DIeSEL} summarizes the distributed operation of our
proposed method. Results presented in~\cite{BeckEldar2013} guarantee
convergence of the algorithm to a stationary point.

\section{Simulation results}
\label{sec:prel-simul-results}

To assess the quality of the algorithm we compared it with a
centralized Extended Kalman Filter (EKF) and with a static
optimization-based distributed algorithm, which only uses noisy ranges
and anchor positions. The EKF parameters were tuned for every
trajectory. For DIeSEL, we assigned the same time window~$T_{0}=5$ for all
trajectories, unless otherwise noted. The static range-only method has
no parameters to tune.

\begin{figure}[!t]
\centering
  \includegraphics[width=\linewidth]{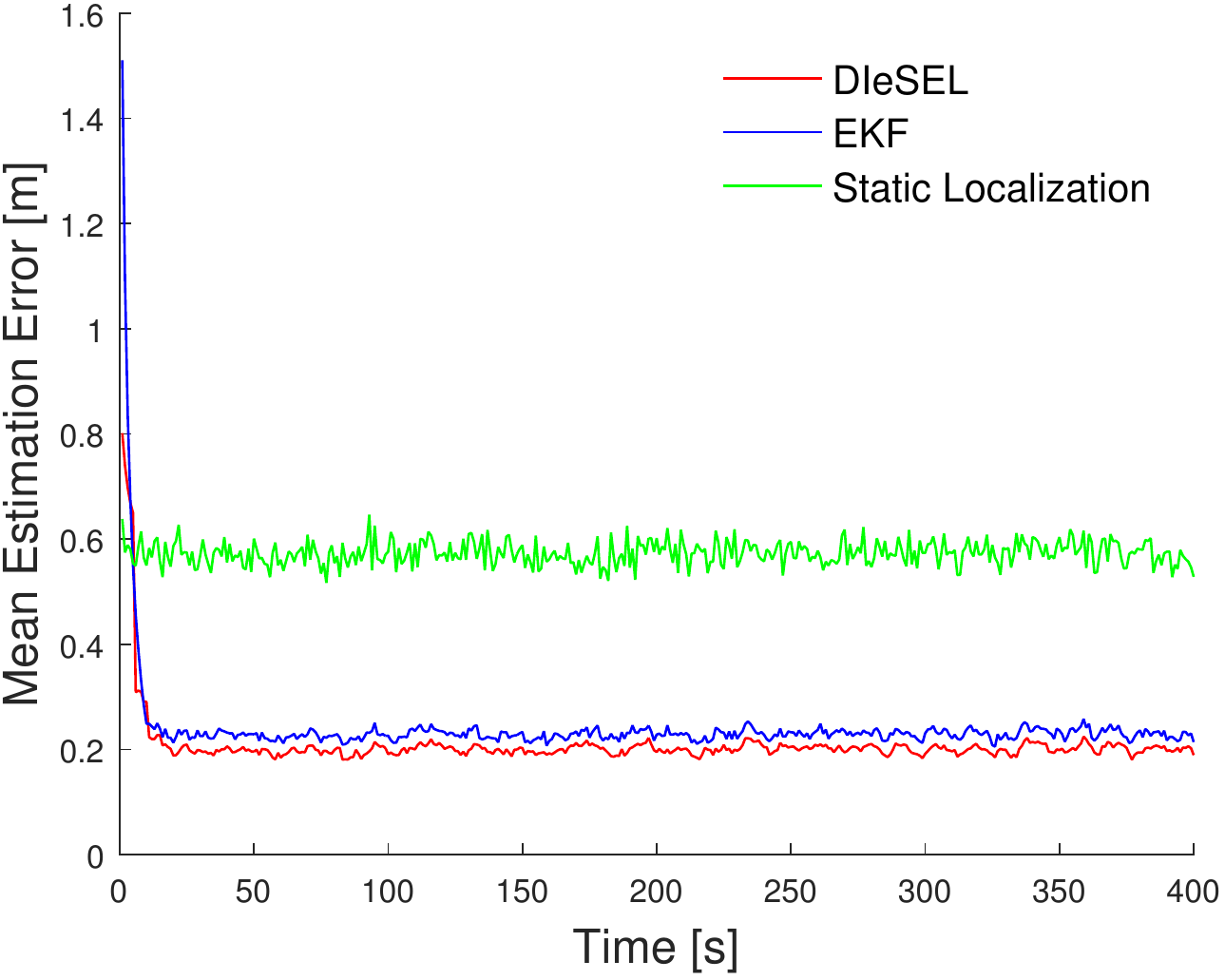}
  \caption{Mean error throughout the lap trajectory. The
    static range-only localization method has the worst performance of
    the benchmark, motivated by the meagre data requirements, and by
    not considering motion in time. The EKF has a larger transient,
    and after 20 sec of the operation's
    onset, DIeSEL's error is already below the plateau of the EKF
    error of estimation, even when well tuned.}
\label{fig:02_laps_mean_est_errors}
\end{figure}

\begin{figure}[!t]
\centering
  \includegraphics[width=\linewidth]{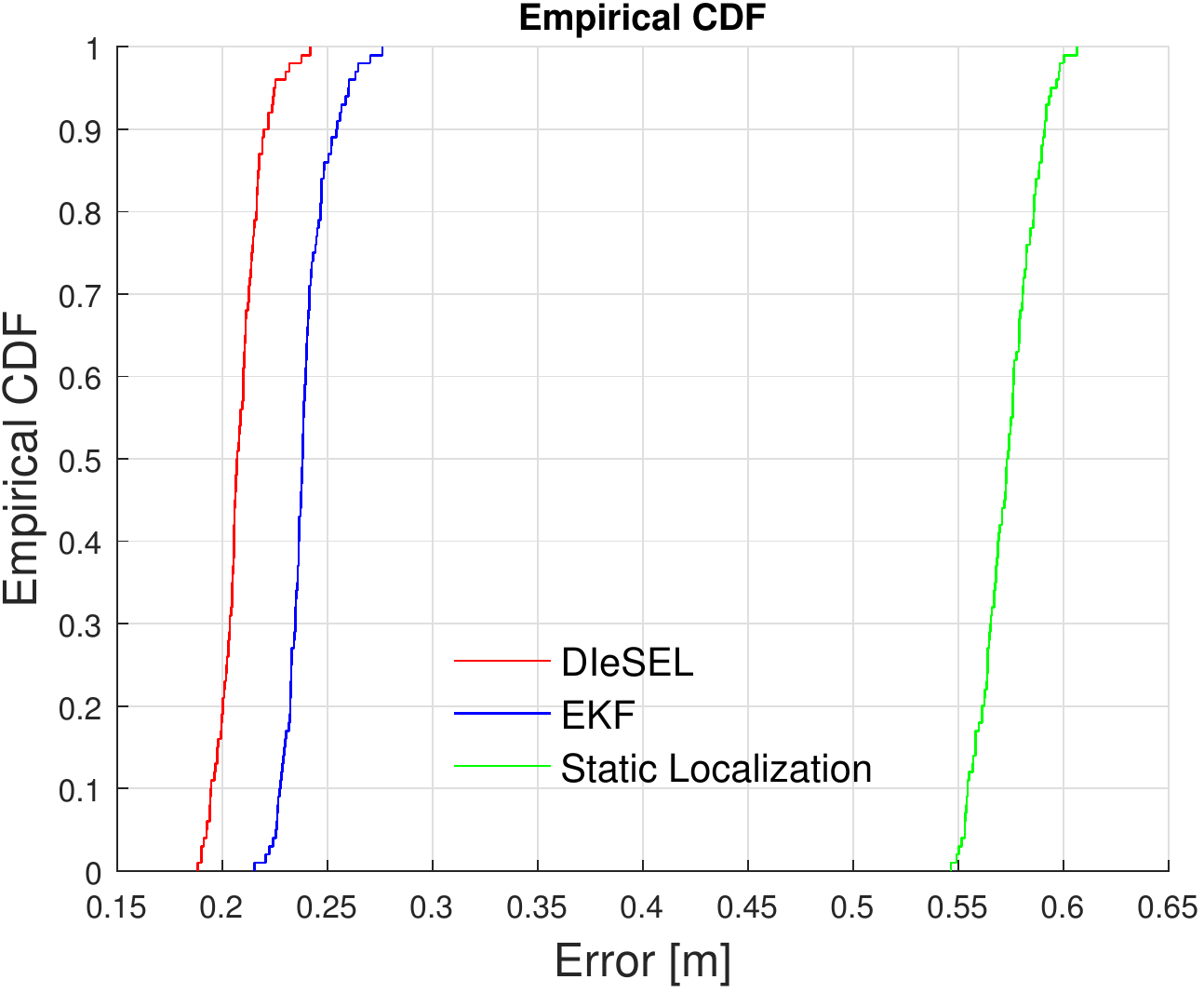}
  \caption{Empirical CDF for the lap experiment with 100 Monte Carlo
    trials. The error is averaged per vehicle and per number of
    trajectory points. Using fewer parameters, the distributed DIeSEL
    algorithm outperforms the centralized EKF. Both outperform the
    optimization based static range-only localization.}
\label{fig:03_laps_cdf}
\end{figure}

\begin{figure}[!t]
\centering
  \includegraphics[width=\linewidth]{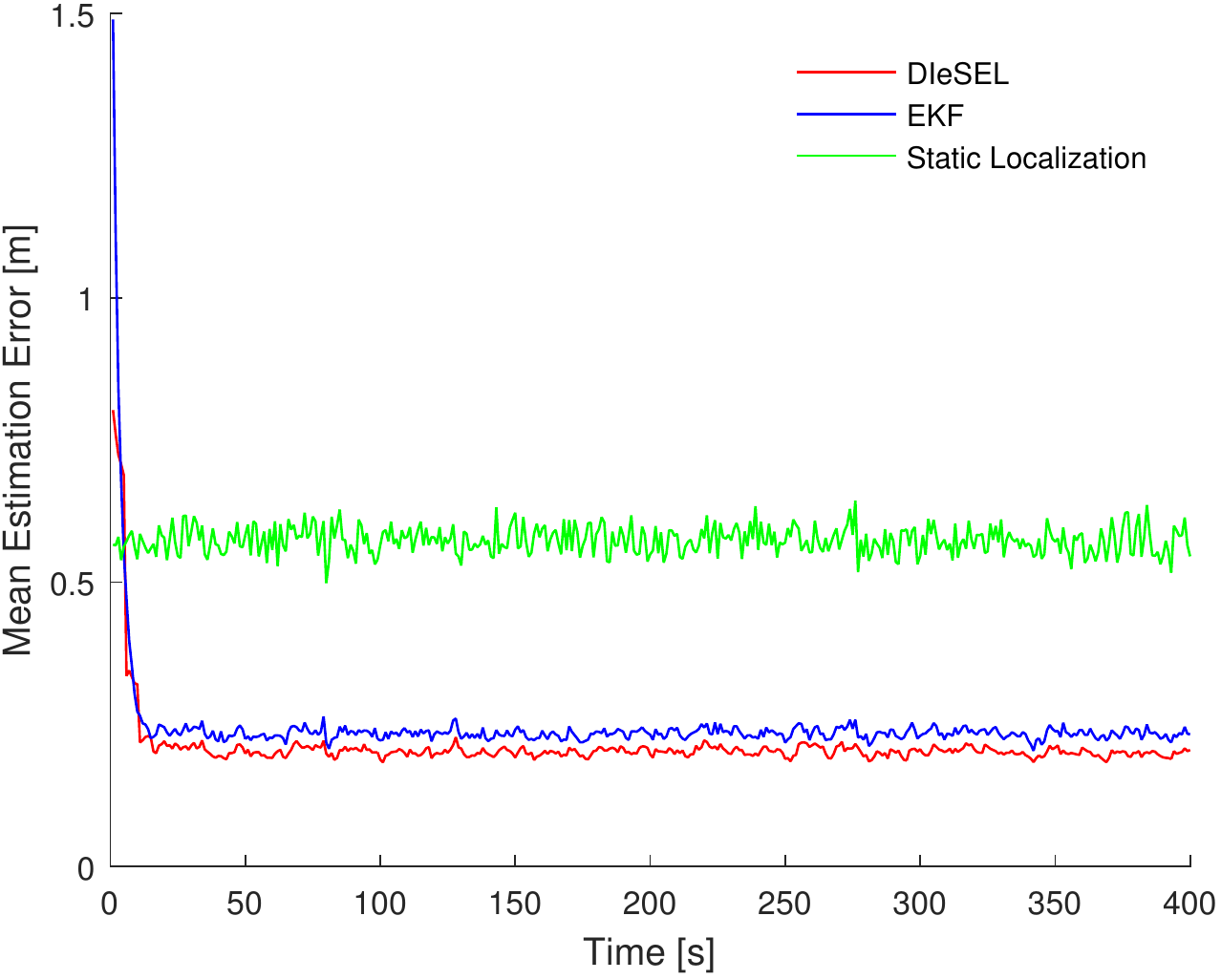}
  \caption{Mean error throughout the lawnmowing trajectory, averaged
    through 100 Monte Carlo trials. The static range-only localization
    method has, still, the worst performance of the benchmark. DIeSEL
    has the steepest response and the best mean estimation error.}
\label{fig:06_lawn_mean_est_errors}
\end{figure}

\begin{figure}[!t]
\centering
  \includegraphics[width=\linewidth]{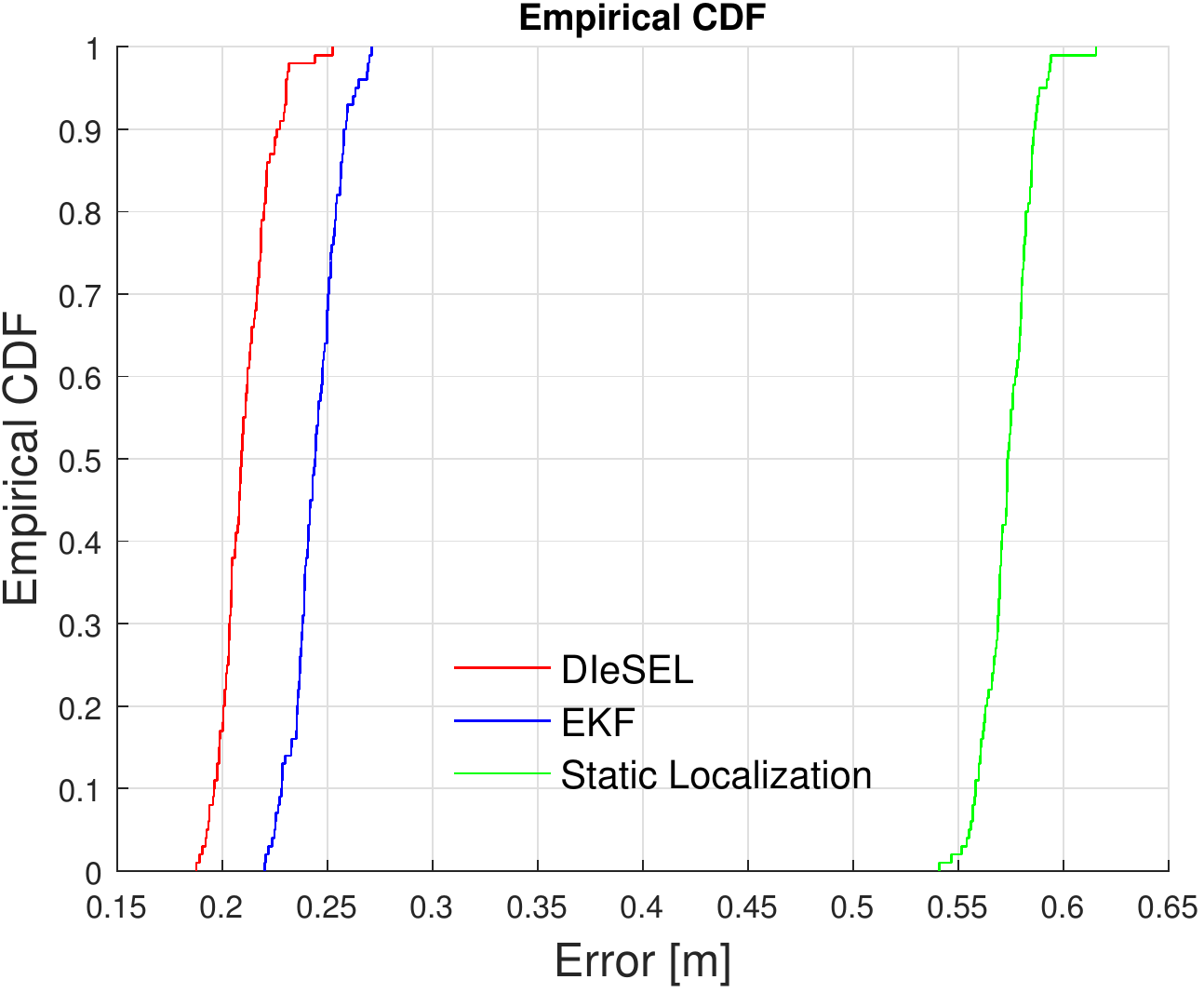}
  \caption{Empirical CDF for 100 Monte Carlo trials on the lawnmowing
    trajectory. DIeSEL is the more precise algorithm, while static
    range-only localization fares the worst. The centralized EKF, when well tuned,
  can achieve at most the second best result.}
\label{fig:07_lawn_cdf}
\end{figure}

\begin{figure}[!t]
\centering
  \includegraphics[width=\linewidth]{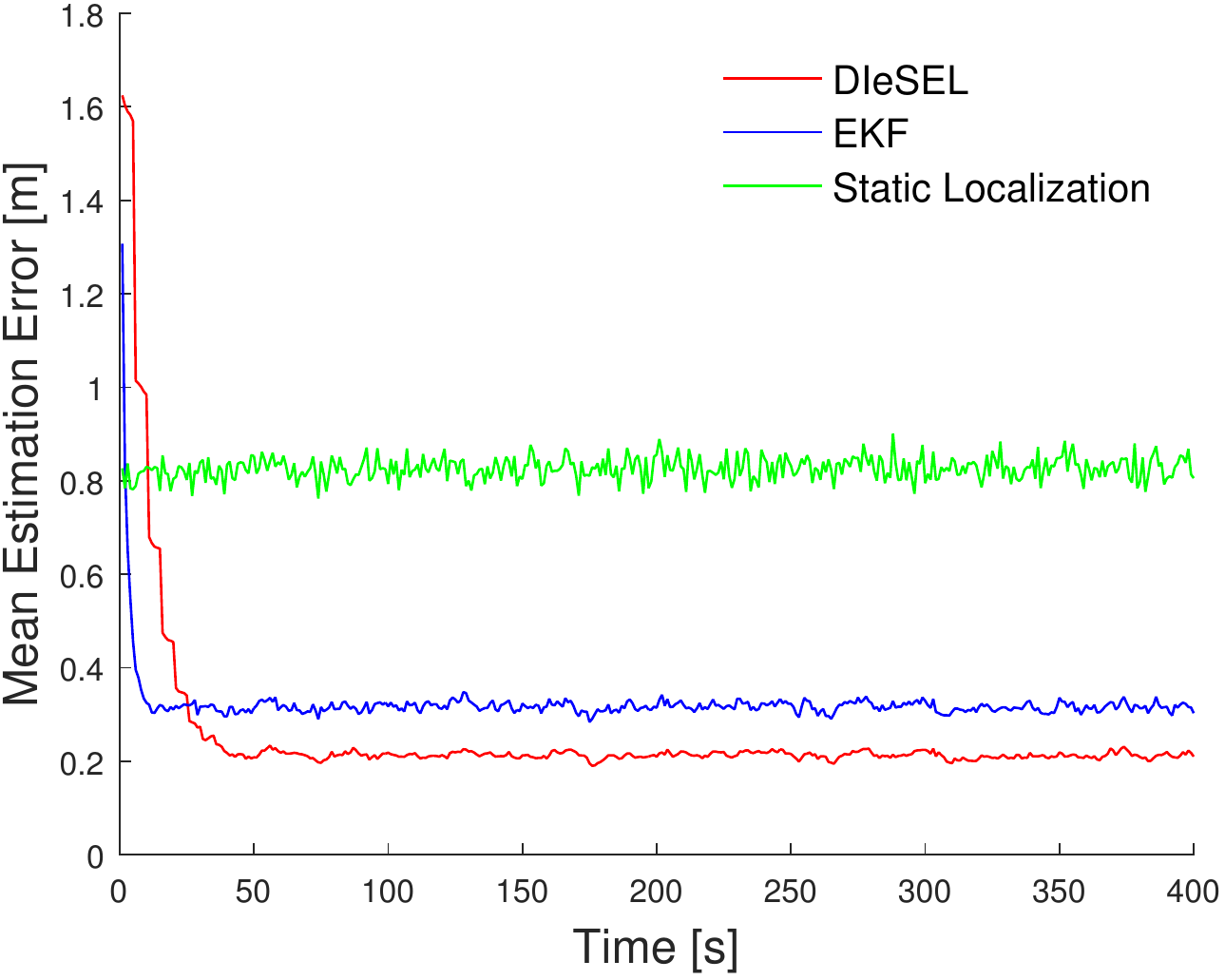}
  \caption{Mean error throughout the helix trajectory. The EKF has a
    steepest response, mainly because the best parameter configuration
    weights much more the model than the measurements, so they are
    considered in a very low proportion in the estimation process. 
    DIeSEL has the best trajectory following error.}
\label{fig:10_helix_mean_est_errors}
\end{figure}

\begin{figure}[!t]
\centering
  \includegraphics[width=\linewidth]{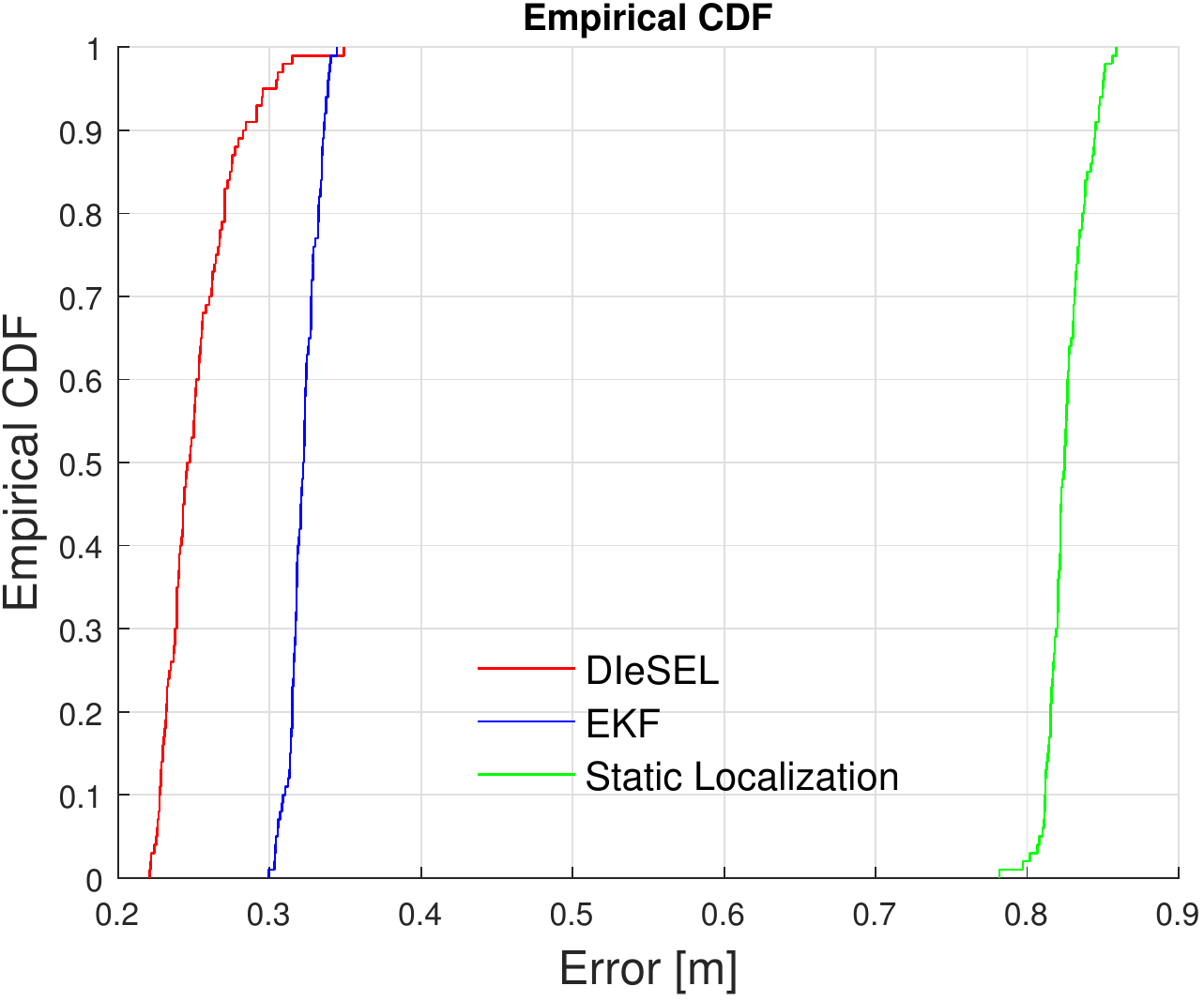}
  \caption{Empirical CDF for the helix trajectory showing the averaged
  error on the number of vehicles and trajectory length. The
  distributed algorithm DIeSEL outperforms both the centralized EKF
  and static localization.}
\label{fig:11_helix_cdf}
\end{figure}

We ran numerical simulations with a four-vehicle team where the center
two vehicles are GPS-enabled (i.e., anchors). The other two navigate only
with noisy range measurements and inertial velocities. 
\begin{figure}[t]
    \centering
    \begin{subfigure}[b]{0.3\textwidth}
        \includegraphics[width=\textwidth]{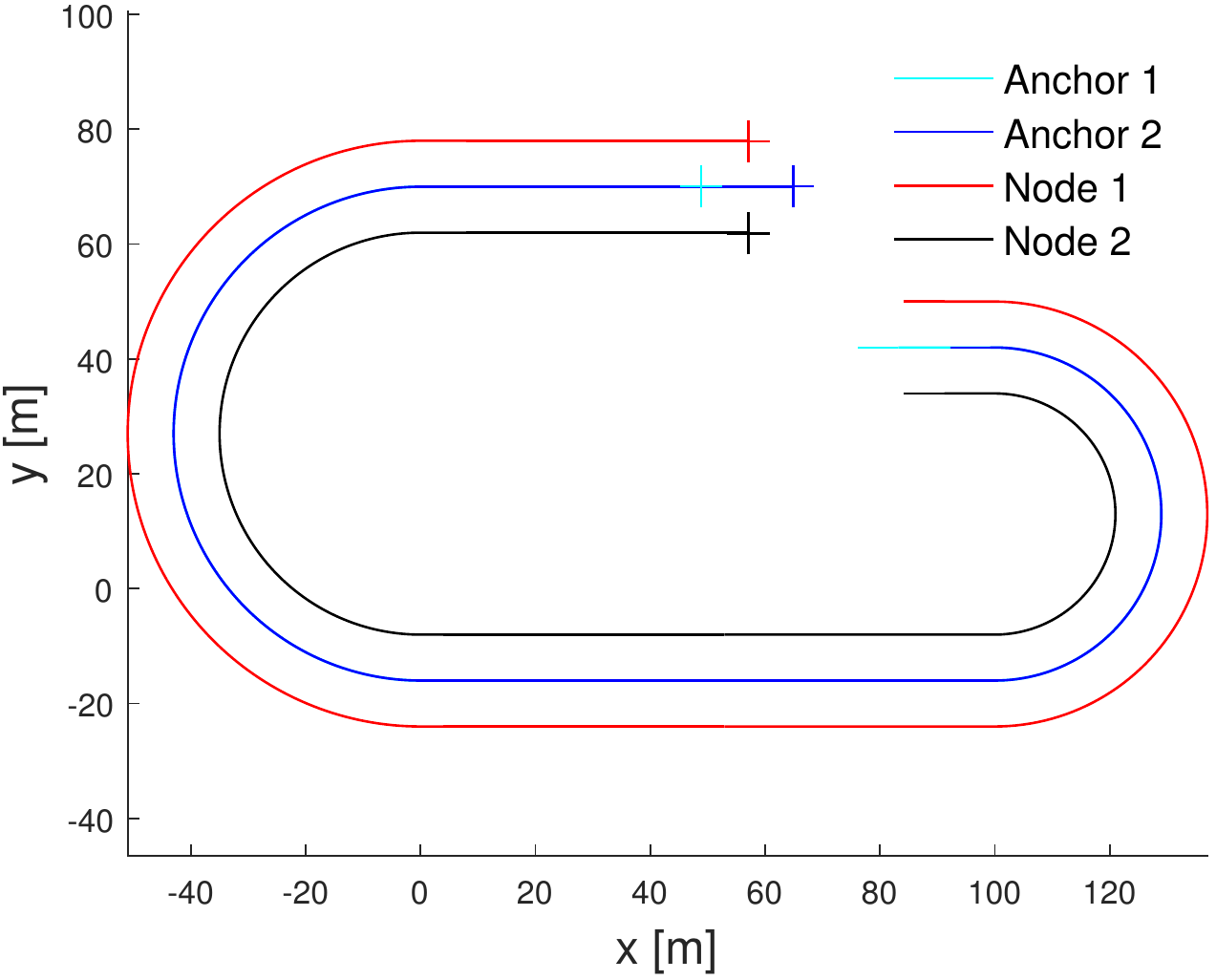}
        \caption{Lap trajectory, with four vehicles (distances in meters).}
        \label{fig:01_laps_traj_exact}
    \end{subfigure}
    ~ 
    \begin{subfigure}[b]{0.3\textwidth}
        \includegraphics[width=\textwidth]{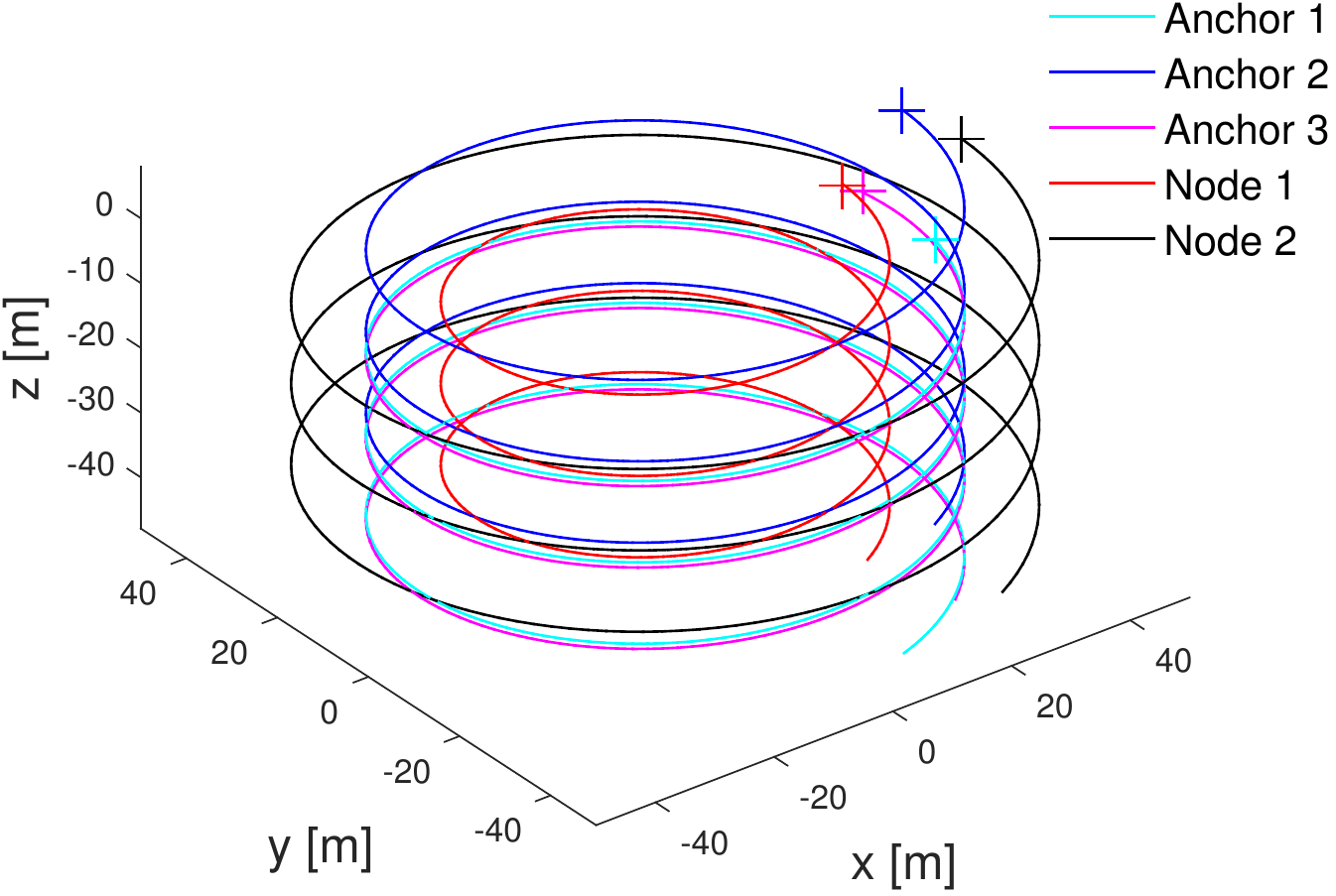}
        \caption{Descending helix.}
        \label{fig:09_helix_traj_exact}
    \end{subfigure}
    ~
    \begin{subfigure}[b]{0.3\textwidth}
        \includegraphics[width=\textwidth]{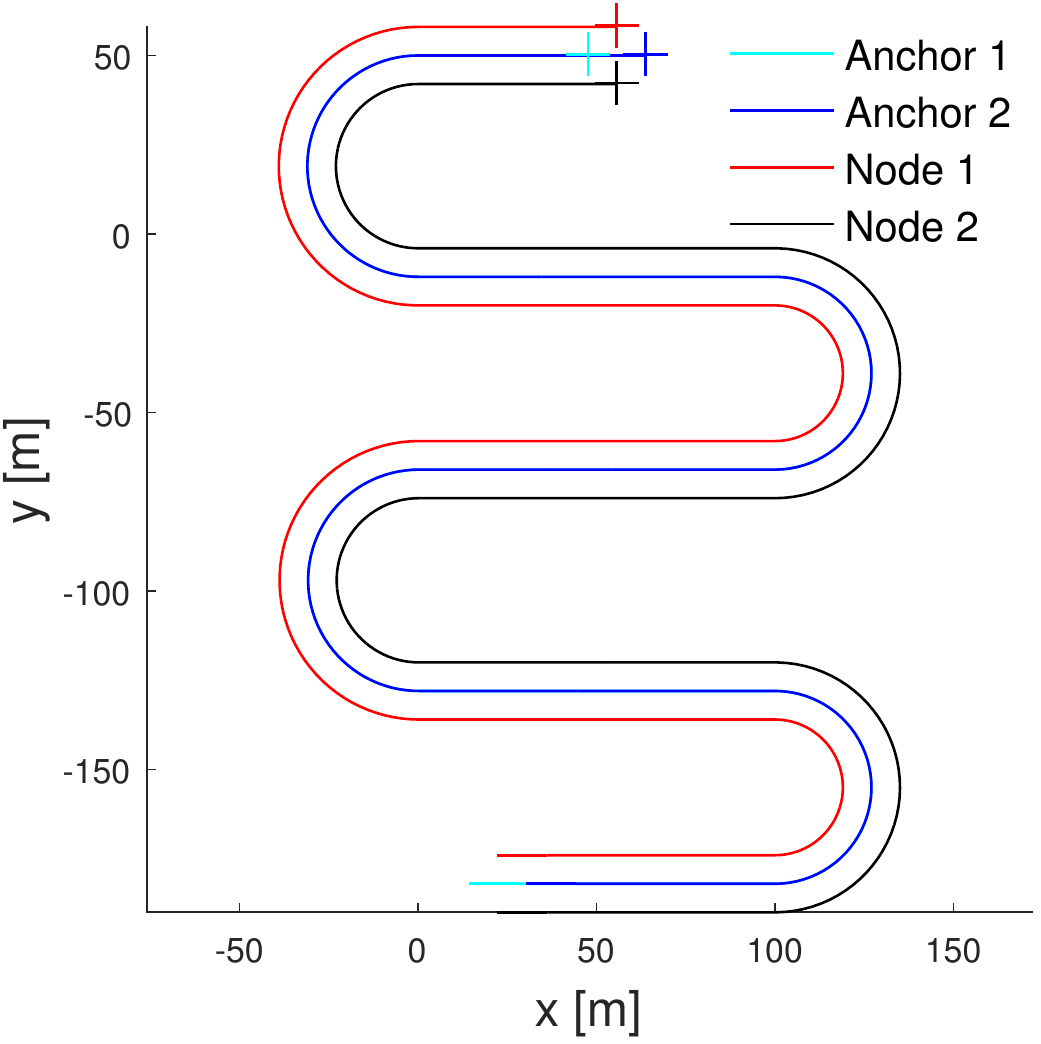}
        \caption{The lawn mower trajectory.}
        \label{fig:05_lawn_traj_exact}
    \end{subfigure}
    \caption{Trajectories for experimental evaluation. Each starting
      position is marked with a plus sign.}\label{fig:trajectories}
\end{figure}
The test trajectories can be seen in
Figure~\ref{fig:trajectories}. Range noise is Gaussian with zero mean
and standard deviation of~0.5m; speed is also contaminated with zero
mean Gaussian noise with standard deviation of 0.01m/s.  We used
trajectories that could span a wide variety of AUV maneuvers: the
lawnmowing, the lap and the 3D helix. Both the EKF and DIeSEL are
initialized with a random position around the true initial point with
a standard deviation of 2m.
Figure~\ref{fig:02_laps_mean_est_errors} depicts the mean estimation
error along the lap trajectory. We can observe that the DIeSEL error
has a smaller transient and follows the trajectory with smaller value
than the centralized EKF. Also, we can see that static range-only
localization fares worse; this phenomenon is not unexpected: static
range-only localization does not take advantage of the information of
velocities and motion continuity.
The normalized error per vehicle, per number of points in the
trajectory is also quite revealing: Figure~\ref{fig:03_laps_cdf}
shows that, albeit using less parameters, and being distributed,
DIeSEL outperforms the EKF. In agreement with the plot
in Figure~\ref{fig:02_laps_mean_est_errors}, the comparison of empirical CDFs
of the velocity- and motion-aware EKF and DIeSEL, and the static
range-only localization shows the advantage of incorporating this
available information.
We performed the same numerical analysis for the lawnmowing
trajectory. The mean estimation error depicted in
Figure~\ref{fig:06_lawn_mean_est_errors} confirms the findings for the
lap. The empirical CDF in Figure~\ref{fig:07_lawn_cdf} agrees with the
corresponding results in the lap trials.
The results for the 3D helix also support the 2D results of our DIeSEL
algorithm: Figure~\ref{fig:10_helix_mean_est_errors} shows a better
tracking error, and the empirical CDF for 100 Monte Carlo trials in
Figure~\ref{fig:11_helix_cdf} exhibits a more remarkable advantage of
DIeSEL in terms of precision, when compared with the centralized EKF
and static range-only localization.

\section{Final remarks}
\label{sec:final-remarks}

DIeSEL aims at precisely --- and distributedly --- localizing a moving
network of agents in a GPS-denied environment, using a generic
kinematic motion model, less dependent on the specific hardware
available than the usual filtering approaches. The proposed algorithm
optimizes the maximum likelihood estimation problem considering noisy
Gaussian range measurements, providing a stable and convergent
sequence of iterates.  DIeSEL's performance in numerical trials
indicates that the performance is resilient to the choice of
trajectory and has a fast transient; also the tracking error is
smaller than a finely tuned centralized EKF, whereas DIeSEL worked out
of the box with a small time window. With no other parameter to tune
than a time window, the proposed DIeSEL algorithm finds the most
precise position of agents in a cooperative network in the set of
benchmark algorithms.

\section*{Acknowledgment}

This research was partially supported by EU-H2020 WiMUST project
(grant agreement No. 645141) and Funda\c{c}\~{a}o para a Cieˆncia e
Tecnologia (project UID/EEA/50009/2013).




\bibliographystyle{IEEEtran}
\bibliography{IEEEabrv,biblos}

\begin{thebibliography}{10}
\providecommand{\url}[1]{#1}
\csname url@samestyle\endcsname
\providecommand{\newblock}{\relax}
\providecommand{\bibinfo}[2]{#2}
\providecommand{\BIBentrySTDinterwordspacing}{\spaceskip=0pt\relax}
\providecommand{\BIBentryALTinterwordstretchfactor}{4}
\providecommand{\BIBentryALTinterwordspacing}{\spaceskip=\fontdimen2\font plus
\BIBentryALTinterwordstretchfactor\fontdimen3\font minus
  \fontdimen4\font\relax}
\providecommand{\BIBforeignlanguage}[2]{{%
\expandafter\ifx\csname l@#1\endcsname\relax
\typeout{** WARNING: IEEEtran.bst: No hyphenation pattern has been}%
\typeout{** loaded for the language `#1'. Using the pattern for}%
\typeout{** the default language instead.}%
\else
\language=\csname l@#1\endcsname
\fi
#2}}
\providecommand{\BIBdecl}{\relax}
\BIBdecl

\bibitem{al2015widely}
H.~Al-Khatib, G.~Antonelli, A.~Caffaz, A.~Caiti, G.~Casalino, I.~B. de~Jong,
  H.~Duarte, G.~Indiveri, S.~Jesus, K.~Kebkal \emph{et~al.}, ``The widely
  scalable mobile underwater sonar technology {(WiMUST)} project: an
  overview,'' in \emph{{OCEANS} 2015-Genova}.\hskip 1em plus 0.5em minus
  0.4em\relax IEEE, 2015, pp. 1--5.

\bibitem{PinheiroMorenoSousaRodriguez2016}
B.~C. Pinheiro, U.~F. Moreno, J.~T.~B. de~Sousa, and O.~C. Rodríguez,
  ``Kernel-function-based models for acoustic localization of underwater
  vehicles,'' \emph{IEEE Journal of Oceanic Engineering}, vol.~PP, no.~99, pp.
  1--16, 2016.

\bibitem{RadWaterschootToonLeus2011}
H.~J. Rad, T.~Van~Waterschoot, and G.~Leus, ``Cooperative localization using
  efficient kalman filtering for mobile wireless sensor networks,'' in
  \emph{Signal Processing Conference, 2011 19th European}.\hskip 1em plus 0.5em
  minus 0.4em\relax IEEE, 2011, pp. 1984--1988.

\bibitem{ShangRumiZhangFromherz2004}
Y.~Shang, W.~Rumi, Y.~Zhang, and M.~Fromherz, ``Localization from connectivity
  in sensor networks,'' \emph{Parallel and Distributed Systems, IEEE
  Transactions on}, vol.~15, no.~11, pp. 961 -- 974, Nov. 2004.

\bibitem{BiswasLiangTohYeWang2006}
P.~Biswas, T.-C. Liang, K.-C. Toh, Y.~Ye, and T.-C. Wang, ``Semidefinite
  programming approaches for sensor network localization with noisy distance
  measurements,'' \emph{Automation Science and Engineering, IEEE Transactions
  on}, vol.~3, no.~4, pp. 360 --371, Oct. 2006.

\bibitem{OguzGomesXavierOliveira2011}
P.~O\u{g}uz-Ekim, J.~Gomes, J.~Xavier, and P.~Oliveira, ``Robust localization
  of nodes and time-recursive tracking in sensor networks using noisy range
  measurements,'' \emph{Signal Processing, IEEE Transactions on}, vol.~59,
  no.~8, pp. 3930 --3942, Aug. 2011.

\bibitem{CalafioreCarloneWei2010}
G.~Calafiore, L.~Carlone, and M.~Wei, ``Distributed optimization techniques for
  range localization in networked systems,'' in \emph{Decision and Control
  (CDC), 2010 49th IEEE Conference on}, Dec. 2010, pp. 2221--2226.

\bibitem{SoaresXavierGomes2014a}
\BIBentryALTinterwordspacing
C.~Soares, J.~Xavier, and J.~Gomes, ``Distributed, simple and stable network
  localization,'' in \emph{Signal and Information Processing (GlobalSIP), 2014
  IEEE Global Conference on}, Dec 2014, pp. 764--768. [Online]. Available:
  \url{http://bit.ly/SimpleStable14}
\BIBentrySTDinterwordspacing

\bibitem{SchlupkothenDartmannAscheid2015}
S.~Schlupkothen, G.~Dartmann, and G.~Ascheid, ``A novel low-complexity
  numerical localization method for dynamic wireless sensor networks,''
  \emph{IEEE Transactions on Signal Processing}, vol.~63, no.~15, pp.
  4102--4114, Aug 2015.

\bibitem{SoaresGomesFerreiraCosteira2017}
\BIBentryALTinterwordspacing
C.~Soares, J.~Gomes, B.~Ferreira, and J.~P. Costeira, ``{LocDyn}: Robust
  distributed localization for mobile underwater networks,'' 2017, submitted.
  [Online]. Available: \url{http://bit.ly/LocDynCS}
\BIBentrySTDinterwordspacing

\bibitem{BeckEldar2013}
\BIBentryALTinterwordspacing
A.~Beck and Y.~Eldar, ``Sparsity constrained nonlinear optimization: Optimality
  conditions and algorithms,'' \emph{SIAM Journal on Optimization}, vol.~23,
  no.~3, pp. 1480--1509, 2013. [Online]. Available:
  \url{http://dx.doi.org/10.1137/120869778}
\BIBentrySTDinterwordspacing

\end{thebibliography}
%



\end{document}